\documentclass[twocolumn,aps,prl,showpacs]{revtex4}
\usepackage{graphicx}
\usepackage{bm}
\begin{document}

\title{Diffusion of heat, energy, momentum and mass in one-dimensional
systems}
\author{Shunda Chen}
\author{Yong Zhang}
\author{Jiao Wang}
\author{Hong Zhao}
\email{zhaoh@xmu.edu.cn}
\affiliation{Department of Physics and Institute of Theoretical Physics
and Astrophysics, Xiamen University, Xiamen 361005, Fujian, China}

\begin{abstract}
We study diffusion processes of local fluctuations of heat, energy,
momentum, and mass in three paradigmatic one-dimensional systems.
For each system, diffusion processes of four physical quantities are
simulated and the cross correlations between them are investigated.
We find that, in all three systems, diffusion processes of energy
and mass can be perfectly expressed as a linear combination of those
of heat and momentum, suggesting that diffusion processes of heat
and momentum may represent the heat mode and the sound mode in the
hydrodynamic theory. In addition, the dynamic structure factor, which
describes the diffusion behavior of local mass density fluctuations,
is in general insufficient for probing diffusion processes of other
quantities because in some cases there is no correlation between them.
We also find that the diffusion behavior of heat can be qualitatively
different from that of energy, and, as a result, previous studies
trying to relate heat conduction to energy diffusion should be
revisited.
\end{abstract}

\maketitle

\section{I. Introduction}

In recent years, stimulated by the rapid progress in both
theoretical and experimental studies, the nonequilibrium transport
properties in low-dimensional systems have become a favored research
topic. In principle, based on the linear response theory
\cite{1KuboSP}, these properties can be revealed by studying the
evolution of spontaneous fluctuations in equilibrium systems. However,
in general how a spontaneous fluctuation may evolve is
still an open question. For example, if a local fluctuation will
simply relax until it vanishes or spread (diffuse) into other parts of
the system, and if fluctuations of different physical quantities
may evolve in the same way, are unclear yet.

To study relaxation processes is of fundamental importance. The
conventional hydrodynamic theory predicts that a perturbation may
induce a heat mode and a sound mode \cite{2TheorySP}, hence the
relaxation of a fluctuation may be understood as a combination of
such two types of hydrodynamic modes. However, in recent decades,
it has been found that the linearized hydrodynamic description may
be invalid in low-dimension systems \cite{LepriEPJB05, Bishop1981,
Bishop1982}. Therefore, it is necessary to investigate, by direct
simulations, the particular properties of the hydrodynamic modes
and show how they manifest themselves in different systems. This
kind of first-hand information can be very helpful for checking
the deviations from the hydrodynamic transport theories and shed
light on how to improve them.

Numerical simulations have special advantages for this aim,
because they are applicable to a large variety of relaxation
processes, many of which are not accessible by present laboratory
experiments. Indeed, among various quantities, at present only the
evolving process of the mass density fluctuations can be measured
in laboratories by inelastic neutron or x-ray scattering experiments
in terms of the dynamic structure factor. The dynamic structure
factor is defined as the Fourier transform of a spatiotemporal
correlation function of local mass density fluctuations
\cite{2TheorySP}. It can be applied to probe the information of
the interparticle interactions and their time evolution, and thus
has been widely studied via theoretical, experimental, and numerical
methods \cite{2TheorySP, D3, D4, D5, D13}. Therefore, verifying the
existence of correlation between the relaxation processes of a given
physical quantity and that of the local mass density fluctuations
has practical importance as well. If they are correlated and the
correlation is made clear by simulations, then with the existing
experimental techniques the relaxation behavior of the given
quantity can be obtained by measuring the dynamic structure factor.
If there is no correlation, then numerical simulations would be
the main tool to explore the former.

Another instance requiring us to clarify if there is any correlation
between relaxation processes of different quantities is encountered
in the study of heat conduction in low-dimensional systems
\cite{LepriPhyRep03, DharAdvPhy08}. Heat conduction is closely
related to the heat relaxation behavior. It is known that the heat
current and the energy current are conceptually different
\cite{Narayan02, Narayan09}, but they may have the same value
at nonequilibrium stationary states \cite{LepriPhyRep03, note1}.
In the literature \cite{Denisov03, LiandWang03, Ewang, E22, E23},
sometimes heat relaxation has been assumed, implicitly, to be the
same as energy relaxation, and the heat conduction properties are
thus deduced based on energy relaxation instead. Given this, to
clarify if heat and energy follow the same relaxation law is
a necessary task.

This work is an effort towards filling these gaps. We shall focus
on the evolution of local fluctuations of energy, heat, momentum,
and mass, and pay particular attention to their correlations. We
shall consider three typical one dimensional (1D) systems as
examples to show the system-dependent relaxation properties. The
rest of this paper is organized as follows: The models to be
studied will be described in the next section, and the method we
use to probe the evolution of local fluctuations will be detailed
in Sec. III. The main results will be provided and discussed
in Sec. IV, followed by a brief summary in the last section.

As we find that most relaxation processes in our study have
characteristics of generalized diffusion--i.e., the corresponding
fluctuations do not decay to zero but spread across the system--
in the following we refer to them as diffusion for the sake of
simplicity. For example, by "diffusion of energy" we mean the
evolution process of local energy fluctuations.

\section{II. Models}

We study three paradigmatic 1D models that have been shown very
useful for exploring the dynamic implications on thermodynamic
properties: one gas model and two lattice models. The gas
model \cite{E22, G40, G41, G42} is a simplified representative of
1D fluids which consists of $N$ hard-core point particles arranged
in order in a 1D box of length $L$ with alternative mass $m_{o}$
for odd-numbered particles and $m_{e}$ for even-numbered particles.
The particles travel freely except for elastic collisions with their
nearest neighbors. The two lattice models are the Fermi-Pasta-Ulam
(FPU) model \cite{FPU43} and the lattice $\phi ^{4}$ model \cite{PH44},
representing lattices with and without the momentum conserving
property, respectively. Their Hamiltonian is $H=\sum_{i}H_{i}$
with $H_{i}=\frac{p_{i}^{2}}{2m_{i}}+\frac{1}{2}(x_{i}-x_{i-1})
^{2}+\frac{1}{4}(x_{i}-x_{i-1})^{4}$ for the FPU model and
$H_{i}=\frac{p_{i}^{2}}{2m_{i}}+\frac{1}{2}(x_{i}-x_{i-1})^{2}+
\frac{1}{4}x_{i}^{4}$ for the lattice $\phi^{4}$ model, where
$H_{i}$, $p_{i}$, $m_{i} $, and $x_{i}$ represent, respectively,
the energy, the momentum, the mass, and the displacement from its
equilibrium position of the $i$th particle.

In our simulations, the system size $L$ is set to be equal to the
particle number $N$, so that the averaged particle-number density
is unity. The local temperature is defined as $T_{i}\equiv \frac
{\langle p_{i}^{2}\rangle }{k_{B}m_{i}}$, where $k_{B}$ (set to
be unity) is the Boltzmann constant and $\langle \cdot \rangle$
stands for the ensemble average. For the 1D gas model, we take
$m_{o}=1$ for odd-numbered particles and $m_{e}=3$ for
even-numbered particles as in Ref. \cite{E22} for the sake of
comparison, and the average energy per particle is unity so that
the system temperature $T=2$. For the FPU model and the lattice
$\phi ^{4}$ model, all particles have a unit mass and the system
temperature is $T=0.5$.

\section{III. Method}

In the equilibrium state, the diffusion behavior of a physical quantity
can be probed by studying the properly rescaled spatiotemporal
correlation function of its density fluctuations \cite{E23, st56,
st57}. The method given in Ref. \cite{E23} will be detailed and
extended to microcanonical systems in the following.

We assume that the systems are microcanonical with periodic
boundary condition and the physical quantity to be considered,
denoted by $\mathcal{A}$, is conserved. The density distribution
function of $\mathcal{A}$ is denoted by $A(x,t)$, where $x$ and
$t$ are the space and the time variables, respectively. In numerical
simulations, in order to calculate the spatiotemporal correlation
function of $A(x,t)$, we have to discretize the space variable.
For this aim we divide the space range of a system into $N_{b}=
\frac{L}{b}$ bins of equal size $b$. The total quantity of $A(x,t)$
in the $j$th bin is denoted by $A_{j}(t)$, defined as $A_{j}(t)
\equiv\int_{x\in j\mathrm {th\, bin}}A(x,t)dx$. The fluctuation
of $A(x,t)$ in the $j$th bin is thus $\Delta A_{j}(t)\equiv A_{j}
(t)-\bar{A}$, where $\bar{A}$ is the ensemble average of $A_{j}(t)$.
These $N_{b}$ bins serve as the coarse-grained space variable.

\begin{figure*}
\vskip-0.5cm \includegraphics[scale=0.8]{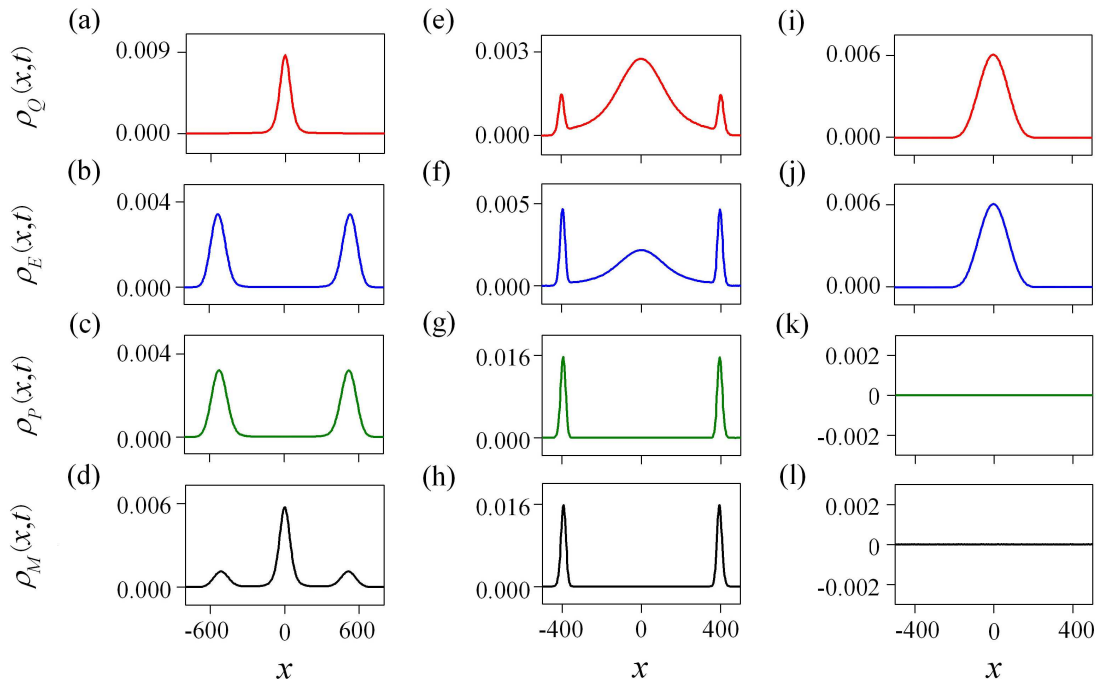}\vskip-0.2cm
\caption{(Color online) The spatiotemporal correlation functions
of heat, energy, momentum and mass, denoted by $\protect\rho_{Q}
(x,t)$, $\protect\rho_{E}(x,t)$, $\protect\rho_{P}(x,t)$, and
$\protect\rho_{M}(x,t)$, respectively, for the 1D gas model (a)-(d),
the FPU model (e)-(h), and the lattice $\protect \phi^{4}$ model
(i)-(l) at $t=300$.}
\end{figure*}

The diffusion characteristics of $\mathcal{A}$ can be captured
by the spatiotemporal correlation function defined as
\begin{equation}
\rho_{A}(\Delta x_{i,j},t)\equiv\frac{\left\langle \Delta A_{j}
(t)\Delta A_{i}(0)\right\rangle }{\left\langle \Delta A_{i}(0)
\Delta A_{i}(0)\right\rangle }-C_{inh},
\end{equation}
where $\Delta x_{i,j}$ denotes the displacement from the $i$th
bin to the $j$th bin, i.e., $\Delta x_{i,j}\equiv (j-i)b$. The
constant $C_{inh}$ represents the inherent correlation resulting
from the fact that $\mathcal{A}$ is conserved, which has nothing
to do with the causal correlation and hence must be deducted
\cite{E23}. For a microcanonical system we have $\sum_{j}\Delta
A_{j}(0)=0$ due to the fact that $\mathcal{A}$ is conserved,
therefore $\sum_{j\ne i}\Delta A_{j}(0)\Delta A_{i}(0)=-\Delta
A_{i}(0) \Delta A_{i}(0)$ and
\begin{equation}
\sum_{j\ne i}\langle \Delta A_{j}(0)\Delta A_{i}(0)\rangle
=-\left\langle \Delta A_{i}(0)\Delta A_{i}(0)\right\rangle.
\end{equation}
As $\langle\Delta A_{j}(0)\Delta A_{i}(0)\rangle$ is statistically
equivalent for all $j\ne i$ due to the homogeneity of space and
time, we have
\begin{equation}
\langle\Delta A_{j}(0)\Delta A_{i}(0)\rangle =-\frac{1}{N_{b}-1}
\langle\Delta A_{i}(0)\Delta A_{i}(0)\rangle.
\end{equation}
At $t=0$, there should be no causal relationship between two
different bins, i.e., $\rho_{A}(\Delta x_{i,j},0)=0$ for $i\ne j$;
we can then obtain that $C_{inh}=-\frac{1}{N_{b}-1}$. Because the
inherent correlation remains unchanged in time, the spatiotemporal
correlation function
\begin{equation}
\rho_{A}(\Delta x_{i,j},t)=\frac{\langle\Delta A_{j}(t)\Delta
A_{i}(0)\rangle}{\langle\Delta A_{i}(0)\Delta A_{i}(0)\rangle}+
\frac{1}{N_{b}-1}
\end{equation}
thus defined can then accurately capture the causal correlation
induced by an initial fluctuation of $\mathcal{A}$ in microcanonical
systems. It is slightly different from the spatiotemporal correlation
function defined in canonical systems, in which the inherent
correlation $C_{inh}$ vanishes \cite{E23}. For the sake of convenience,
in the following the notation $x$ will be used to replace $\Delta
x_{i,j}$ without confusion. Because the spatiotemporal correlation
function defined above gives the causal relation between a local
fluctuation and the effects it induces at another position (with
a displacement $x$) and at a later time (with a time delay $t$),
it is in essence equivalent to the probability density function
that describes the diffusion process of the fluctuation.

It should be noted that the coarse-grained space variable we
have taken is important for obtaining the correct spatiotemporal
correlation function. If the indexes of particles are used as the
space variable, the corresponding correlation function could be
dramatically different, because the indexes do not reflect the
real physical positions of the particles and thus may cause large
position fluctuations \cite{DharCommt}. Indeed, as will be presented
in the next section, the 1D gas model's spatiotemporal
correlation function of energy has a two-peak structure [see
Fig. 1(b)]. But if the indexes of particles are used, it shows
a three-peak structure instead \cite{E22,Denisov11}. We therefore
emphasize that the coarse-grained space variable is essential for
studying the $spatial$ diffusion, which is exactly our aim here.

The 1D gas model is efficiently simulated by using the
event-driven algorithm that employs the heap data structure to
identify the collision times \cite{G42}. For the FPU model and
the lattice $\phi^{4}$ model, a Runge-Kutta algorithm of 7th to 8th
order is adopted for integrating the motion equations, and the
Andersen thermostat \cite{A54} is utilized to thermalize the
system for preparing the equilibrium systems. In calculating
the spatiotemporal correlation functions a periodic boundary
condition is applied and $N=4000$ particles are considered, but
we have checked and verified that for larger $N$ the simulation
results remain the same. For all three models the equilibrium
systems are prepared by evolving the systems for a long enough
time ($>10^{8}$ time units of the models) from properly assigned
random states \cite{Y52}, then the system is evolved in isolation.
The size of the ensemble for averaging is larger than $10^{10}$.

\begin{figure*}
\vskip-.4cm \includegraphics[scale=0.85]{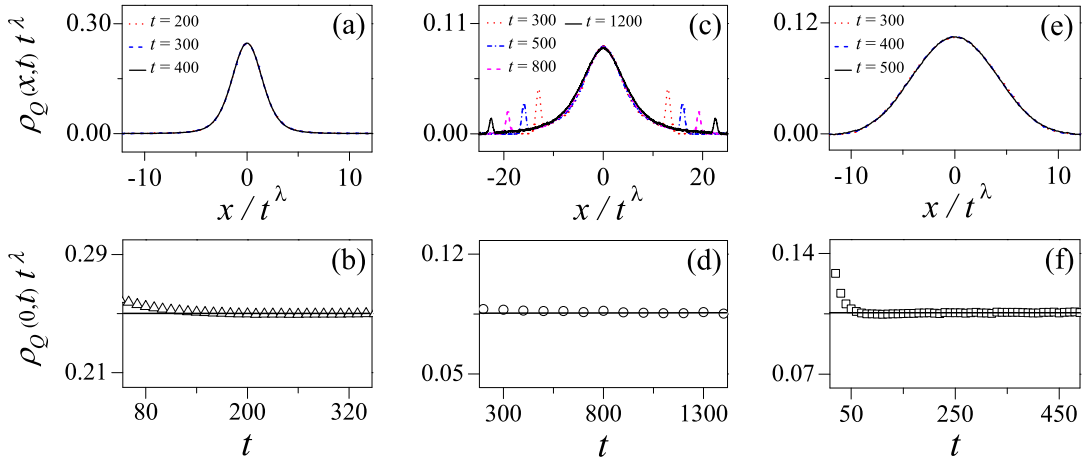} \vskip-.6cm
\caption{(Color online) Rescaled profiles of the spatiotemporal
correlation function of heat $\protect\rho_{Q}(x,t)$ for all
three models. (a)-(b) are for the gas model with rescaling
factor $\protect\lambda=0.59$ obtained via best fitting.
In (a) $\protect\rho_{Q}(x,t)t^{\protect\lambda}$ vs $x/
t^{\lambda}$ at three different times are compared and in (b)
$\protect \rho_{Q}(0,t)t^{\protect\lambda}$ vs time is shown.
(c)-(d) and (e)-(f) are the same as (a)-(b) but for the FPU model
with rescaling factor $\protect\lambda =0.60 $ and the
lattice $\protect\phi^{4}$ model with rescaling factor
$\protect\lambda=0.50$ respectively. As a result of rescaling,
three curves in (a) and (e) overlap and are indistinguishable.
In (c), on each curve there are two side peaks symmetric with
respect to $x=0$, which from the center to the two sides are
at $t=300$, 500, 800, and 1200, respectively.}
\end{figure*}

\section{IV. Results and Discussions}

We probe the diffusion behavior of a given physical quantity by
studying its spatiotemporal correlation function. We are particularly
interested in the diffusion behaviors of heat, energy, momentum, and
mass, whose density functions are respectively denoted by $Q(x,t)$,
$E(x,t)$, $P(x,t)$, and $M(x,t)$, and the corresponding spatiotemporal
correlation functions are denoted by $\rho_{Q}(x,t)$, $\rho_{E}(x,t)$,
$\rho_{P}(x,t)$, and $\rho_{M}(x,t)$. For 1D systems, the heat density
function is defined as $Q(x,t)=E(x,t)-\frac{(\bar{E}+\bar{F})M(x,t)}
{\bar{M}}$ \cite{HE45}, where $\bar{E}$ ($\bar{M}$) and $\bar{F}$ are,
respectively, the spatially averaged energy (mass) density and the
internal pressure of the system in equilibrium state. In our
simulations, the density functions $E(x,t)$, $P(x,t)$, and $M(x,t)$
are numerically measured first, based on which $Q(x,t)$ is obtained
as well. Then the corresponding spatiotemporal correlation functions
are evaluated straightforwardly.

The spatiotemporal correlation functions of all three models
at an example time $t=300$ are shown in Fig. 1. It can be seen that
the diffusion behavior of the same quantity may vary from system to
system (see, e.g., any row in Fig. 1 for a comparison) and in the
same system, different quantities may have dramatically different
diffusion properties as well, though for some of them, such as
energy and momentum in the gas model, the diffusion behaviors could
be the same. Also, it can be found that except for $\rho_{P} (x,t)$ and
$\rho _{M}(x,t)$ for the lattice $\phi ^{4}$ model, all other
spatiotemporal correlation functions conserve their total volume,
i.e., $\int \rho _{A}(x,t)dx=1$, and we can identify either one
standing center peak, or two moving side peaks, or the "superposition"
of them. In every case where two side peaks appear, we find that
the centers of the side peaks move outwards at a constant speed.
More specifically, the side peaks move outwards at a speed $v=1.75$
in the gas model and at $v=1.32$ in the FPU model. In the lattice
$\phi ^{4}$ model where the total momentum is not conserved due
to existence of on-site potentials, we find that $\rho _{P}(x,t)$
and $\rho _{M}(x,t)$ decay exponentially and relax to zero rapidly.
This is the reason why in Figs. 1(k) and 1(l) no structure is
identified.

We find that both $\rho _{M}(x,t)$ and $\rho _{E}(x,t)$ can be
perfectly expressed as a linear combination of $\rho _{Q}(x,t)$
and $\rho_{P}(x,t)$. Our data show that in the gas model $\rho_{M}
(x,t)=\frac{2}{3}\rho_{Q}(x,t)+\frac{1}{3}\rho_{P}(x,t)$ and
$\rho_{E}(x,t)=\rho _{P}(x,t)$, while in the FPU model we have
$\rho _{M}(x,t)=\rho _{P}(x,t)$ and $\rho_{E}(x,t) =0.78 \rho_{Q}
(x,t)+0.22\rho _{P}(x,t)$, and in the lattice $\phi^{4}$ model
$\rho_{E}(x,t)=\rho_{Q}(x,t)$ and $\rho_{M}(x,t)=\rho_{P}(x,t)=0$.

We conjecture that functions $\rho_{Q}(x,t)$ and $\rho_{P}(x,t)$
can be identified with the heat mode and the sound mode, respectively. 
Indeed, the function $\rho _{Q}(x,t)$ represents the heat mode 
because by definition it describes the motion of heat exclusively 
\cite{HE45}. The function $\rho_{P}(x,t)$ describes the collective 
motion carrying the memory of the initial moving directions of 
the particles. In both the gas and the FPU models, $\rho_{P}(x,t)$ 
has a bimodal structure and the peaks move outwards at a constant 
speed, hence can be reasonably related to the sound mode. To make 
our conjecture more convincing, we study the dynamic structure 
factor represented by the function $\rho_{M}(x,t)$. In the gas 
model, we have measured with high precision the volume of the 
center peak (i.e., the area enclosed by the center peak curve 
and the abscissa) of the function $\rho_{M} (x,t)$, finding it to be 
$\frac{2}{3}$, and that of the two side peaks to be $\frac{1}{3}$ 
[see Fig. 1(d)]; the ratio of them equals $2$, which is unchanged 
in time and is in perfect agreement with the Landau-Placzek ratio 
\cite{LP46,LP47} of an ideal gas. In addition, if we multiply 
$\rho _{M}(x,t)$ by a factor of $3$, its side peaks will overlap 
with $\rho _{P}(x,t)$. Similarly, multiplying $\rho _{M}(x,t)$ 
by $\frac{3}{2}$, its center peak will overlap with $\rho _{Q}
(x,t)$. These facts convincingly suggest that $\rho _{Q}(x,t)$ 
and $\rho _{P}(x,t)$ represent the heat mode and the sound mode. 
In the FPU model, there are only two side peaks on $\rho_{M}(x,t)$, 
therefore the ratio of the area of the center peak to that of 
the two side peaks is zero, which is also in agreement with the 
Landau-Placzek ratio of this model. The function $\rho_{M}(x,t)$ 
thus represents only the sound mode. The function $\rho_{P}(x,t)$ 
represents the sound mode as well, because it is identical to 
$\rho_{M}(x,t)$ as shown in Figs. 1(g) and 1(h).

\begin{figure}
\vskip-.2cm
\hskip-.5cm
\includegraphics[scale=0.71]{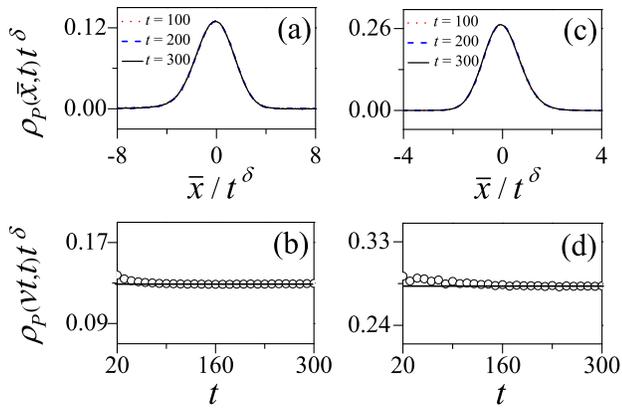}
\vskip-.4cm
\caption{(Color online)  Rescaled profiles of the side peaks on the
spatiotemporal correlation function of momentum $\rho_{P}(x,t)$ for
the gas model and the FPU model. (a) and (b) are for the gas model with
rescaling factor $\delta=0.64$ obtained via best fitting. In
(a) $\rho_{P}(\bar{x},t)t^{\delta}$ vs $\bar{x}/t^{\delta}$ at
three different times are compared, where $\bar{x}=x-vt$ ($v$ is the
speed of the side peak). In (b) $\rho_{P}(vt,t)t^{\delta}$ vs time
is shown. (c) and (d) are the same as (a) and (b) but for the FPU model with
rescaling factor $\delta=0.50$. The fact that three curves in
(a) and (c) overlap and are indistinguishable indicates the perfect
scaling properties of the side peaks of $\rho_{P}(x,t)$.}
\end{figure}

As a consequence, diffusion properties of heat and momentum can
characterize all others diffusion processes. Let us first discuss
diffusion of heat by considering $\rho _{Q}(x,t)$. We find that
there is an interesting scaling property in $\rho _{Q}(x,t)$ in all
three systems, i.e., $\rho _{Q}(x,t)$ is invariant upon rescaling
$x\rightarrow t^{\lambda }x$ so that $t^{\lambda }\rho _{Q}(x,t)=
t_{0}^{\lambda }\rho _{Q}(x_{0},t_{0})$ for $x=(\frac{t}
{t_{0}})^{\lambda }x_{0}$ {[}see Figs. 2(a) and 2(e){]}. For
the gas model and the lattice $\phi ^{4}$ model $\rho _{Q}(x,t)$
is a unimodal function; the simulation results suggest that
$\lambda =0.59$ for the former and $\lambda =0.50$ for the latter.
Neglecting the decaying side peaks, $\rho _{Q}(x,t)$ of the FPU
model has the same scaling invariance property with $\lambda =0.60$
{[}see Fig. 2(c){]}. As $\rho _{Q}(x,t)$ conserves its volume, we
have $\rho _{Q}(x,t)dx=\rho _{Q}(x_{0},t_{0})dx_{0}$, which leads
to the result that the variance of $\rho _{Q}(x,t)$ goes in time as $\langle
x^{2}(t)\rangle=\langle x_{0}^{2}(t_{0})\rangle (\frac{t}
{t_{0}})^{2\lambda}$; i.e., a heat fluctuation diffuses in
a power law $\langle x^{2}(t)\rangle\sim t^{\beta}$ with the
diffusion exponent $\beta=2\lambda$. Thus, we obtain $\beta=
1.18$, $1.20$, and $1.00$ for the gas model, the FPU model, and
the lattice $\phi ^{4}$ model, respectively, indicating that a
heat fluctuation undergoes superdiffusion in the gas model and
the FPU model but normal diffusion in the lattice $\phi ^{4}$ model.
As mentioned above, the center peak of $\rho_{M}(x,t)$ in the gas
model can be rescaled and overlap perfectly with $\rho_{Q}(x,t)$.
So can the center peak of $\rho_{E}(x,t)$ in the FPU model and in
the $\phi^{4}$ model. To summarize, these peaks relax in the same 
manner as that of $\rho_{Q}(x,t)$.

Now let us turn to the sound mode. If $\rho _{A}(x,t)$ has two
side peaks moving at a constant speed $v$ and conserving their
volumes, then $\langle x^{2}(t)\rangle \sim \int_{vt-\Delta x}^
{vt+\Delta x}(vt)^{2}\rho_{A}(x,t)dx\sim t^{2}$, where $\Delta
x$ represents the width of the peaks. Hence all the processes
that involve the sound mode should fall into the class of ballistic
diffusion, including the diffusion processes of energy, momentum,
and mass density in the gas model and in the FPU model. Particularly,
for $\rho _{E}(x,t)$ in the FPU model we have $\langle x^{2}(t)
\rangle \sim at^{1.20}+ct^{2}$ since $\rho_{E}(x,t)=0.78\rho_{Q}
(x,t)+0.22\rho_{P}(x,t)$ as mentioned above, where $a$ and $c$
are constants. Therefore, though the center peak relaxes in a
superdiffusive manner, its asymptotic diffusion behavior will be
dominated by $\langle x^{2}(t)\rangle \sim ct^{2}$ even if $a$
is much larger than $c$.

The peaks of $\rho_{P}(x,t)$ disperse while moves ballistically.
The dispersion reveals the information of the sound attenuation.
Similarly, as for the center peak of $\rho_{Q}(x,t)$, we find
that there is also an interesting scaling property of the side
peaks of $\rho_{P}(x,t)$, i.e., the side peak, taking the right
one for example, of $\rho_{P}(x,t)$ is invariant upon rescaling
$\bar{x}\rightarrow t^{\delta}\bar{x}$, where $\bar{x}=x-vt$
($v$ is the speed of the side peak), so that $t^{\delta}\rho_{P}
(\bar{x},t)=t_{0}^{\delta}\rho_{P}(\bar{x}_{0},t_{0})$ for $\bar{x}=(\frac{t}{t_{0}})^{\delta}\bar{x}_{0}$, where
$\bar{x}_{0}= x-vt_{0}$. As shown in Fig. 3 for the gas
model and the FPU model, the simulation results suggest that
$\delta=0.64$ for the former and $\delta=0.50$ for the latter.

The above results have some important implications. For example,
they indicate that the dynamic structure factor is not sufficient
to capture all diffusion processes. In the gas model, as $\rho_{M}
(x,t)=\frac{2}{3}\rho_{Q}(x,t)+\frac{1}{3}\rho _{P}(x,t)$ and
$\rho_{E}(x,t)=\rho _{P}(x,t)$, indeed energy and momentum diffusion 
can be revealed by the side peaks of $\rho_{M}(x,t)$ and heat 
diffusion can be extracted from the center peak of $\rho _{M}(x,t)$. 
But in the FPU model, though the dynamic structure factor may
characterize the momentum diffusion because $\rho _{M}(x,t)=\rho
_{P}(x,t)$, it is useless for exploring energy and heat diffusion.
In the lattice $\phi ^{4}$ model, the function $\rho _{M}(x,t)$
does not give any interesting information {[}See Fig. 1(l){]}.
Therefore, it is necessary to investigate the diffusion behavior
of other physical quantities case by case, not only when they have
no correlation with the diffusion behavior of mass density so that
they can not be probed by the dynamic structure factor, but also
when the correlation exists but we want to ascertain how to extract
the relaxation properties of other quantities from the dynamic
structure factor.

\begin{figure*}
\vskip-.3cm \includegraphics[scale=0.78]{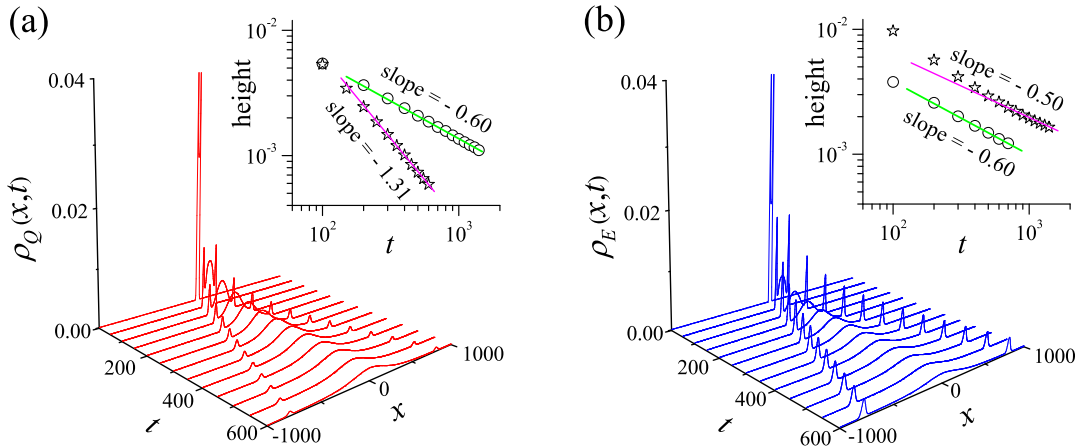}\vskip-.3cm
\caption{(Color online) (a) The spatiotemporal correlation
function of heat, $\protect\rho_{Q}(x,t)$, for the FPU model.
The inset shows the log-log plot of time dependence of the
heights of the center peak (open circles) and the side peaks
(open stars). (b) The same as (a) but for the spatiotemporal
correlation function of energy, $\protect\rho_{E}(x,t)$, of
the FPU model.}
\end{figure*}

Another important implication is that there is no definite
correlation between diffusion of heat and energy. The results
presented in Fig. 1 also suggest that, in a system, diffusion
behavior of energy can be completely different from that of heat
and hence may not provide any information of the latter. For
example, in the gas model [see Figs. 1(b) and 1(a)], there
are two moving peaks in the spatiotemporal correlation function
of energy $\rho _{E}(x,t)$, but there is only one standing peak
in that of heat $\rho _{Q}(x,t)$. In the FPU model, though at
an early stage both $\rho _{E}(x,t)$ and $\rho _{Q}(x,t)$ have
a three-peak structure [see Figs. 1(f) and 1(e)], there is
a significant difference between them: while the former keeps
its three-peak structure throughout, the two side peaks in the
latter keep shrinking and asymptotically $\rho _{Q}(x,t)$ becomes
unimodal. To scrutinize this difference, we compare in Fig. 4 the
time evolution of $\rho _{Q}(x,t)$ and $\rho _{E}(x,t)$. We find
in both of them the half-height width of the side peaks widens as
$l\sim t^{0.50}$, but their height decays as $h\sim t^{-1.31}$
in $\rho _{Q}(x,t)$ rather than $h\sim t^{-0.50}$ as in
$\rho _{E}(x,t)$. As a result the side peaks of $\rho _{E}(x,t)$
keep their volumes unchanged since $lh$ is time independent, in
clear contrast to those of $\rho _{Q}(x,t)$ that keep shrinking 
as $lh\sim t^{-0.81}$. Hence over time $\rho _{Q}(x,t)$ and 
$\rho _{E}(x,t)$ will become qualitatively different. Of all 
three models we find that only in the lattice $\phi ^{4}$ model 
are the diffusion behaviors of energy and heat the same {[}see 
Figs. 1(i) and 1(j){]}.

These results suggest that the previous studies trying to
establish a universal connection between {\it energy} diffusion
and {\it heat} conduction \cite {E22,E23} should be revisited.
Conceptually, it is heat diffusion, rather than energy diffusion,
that should and can be meaningfully related to heat conduction.
For 1D momentum conserving systems, it has been found that
generally the heat conductivity $\kappa $ diverges in the
thermodynamic limit as $\kappa \sim L^{\alpha }$ with $0.25\leq
\alpha \leq 0.5$ \cite {G42, LepriPhyRep03, DharAdvPhy08, Narayan02,
MC1, Beijeren12, sol1, KN1, Mai07, FA51}, and on the other hand,
energy diffuses in time as $\sim t^{\beta }$. Two formulas,
$\alpha=\beta-1$ \cite{Denisov03} and $\alpha=2-\frac{2}{\beta }$
\cite{LiandWang03}, have been proposed to connect the two
exponents $\alpha$ and $\beta $ universally. But as shown in
our gas model, the energy fluctuations spread ballistically
and thus the diffusion of energy is characterized by $\beta=2$,
such that $\alpha $ and $\beta $ definitely do not agree with
either of the two formulas. In the FPU model, due to the ballistic
behavior of the two side peaks on $\rho_{E}(x,t)$, $\beta$
will asymptotically saturate at $\beta=2$, again disobeying the
two formulas.

From the hydrodynamic point of view, the reason why there is no
universal connection between energy diffusion and heat conduction
is conceptually easy to understand: the former is also affected
by advection, i.e., the sound mode. This has been well shown by
our simulation results that, in all three models, the diffusion
process of energy can be perfectly expressed as a linear combination
of those of heat and momentum.

\section{V. Summary}

The method for calculating the spatiotemporal correlation functions
in microcanonical systems has been shown to be effective in probing
the diffusion processes in equilibrium systems. By this method, the
diffusion processes of local fluctuations of heat, energy, momentum,
and mass in three one-dimensional systems are explored in detail.
It is found that diffusion of energy and mass can be expressed as a
linear combination of $\rho_{Q}(x,t)$ and $\rho_{P}(x,t)$, which
we conjecture to be representatives of the heat mode and the sound
mode, respectively. There is a scaling in function $\rho_{Q}(x,t)$,
i.e., $\rho_{Q}(x_{0},t_{0})=(\frac{t}{t_{0}})^{\lambda}\rho_{Q}
((\frac{t}{t_{0}})^{\lambda}x_{0},t)$, in all three models. For
the lattice $\phi^{4}$ model, $\lambda=0.5$, indicating normal
diffusion. For the gas model and the FPU model, $\lambda=0.59$ and
$0.60$, respectively, indicating superdiffusion. The
function $\rho_{P}(x,t)$ vanishes in the lattice $\phi^{4}$ model
due to the momentum-nonconserving property. For the gas model and
the FPU model, $\rho_{P}(x,t)$ conserves its volume and follows the
scaling relation $\rho_{P}(\overline{x}_{0}, t_{0})=(\frac{t}{t_{0}}) ^{\delta}\rho_{P}((\frac{t}{t_{0}})^{\delta}\overline{x}_{0},t)$
with $v=1.75$, $\delta=0.64$ in the gas model (at temperature 
$T=2$) and $v=1.32$, $\delta=0.50$ in the FPU model.

We have revealed correlations between different diffusion processes.
It is found that the diffusion behaviors of different physical
quantities may be distinctively different, and the correlations
between them could be very complex. The diffusion behavior of a
physical quantity may vary from system to system, hence they should
be studied case by case. Diffusion of heat, energy, and momentum is
correlated with that of mass density fluctuations in the gas model,
which implies that they can be probed by measuring the dynamic
structure factor. In the two lattice models, the dynamic structure
factor provides no information of heat diffusion.

A particular finding is that diffusion of energy can be qualitatively
different from that of heat, hence a universal connection may not
exist between energy diffusion and heat conduction (though we
conjecture that instead of energy diffusion, a universal connection
may be established between {\it heat} diffusion and {\it heat}
conduction). This is different from the relationship between the
energy current and the heat current \cite{Narayan02, Narayan09},
which turns out to be the same \cite{LepriPhyRep03} at
nonequilibrium stationary state.

We have not studied two- and three-dimensional systems. In
one of our studies \cite{Y52} it has been shown that the particle
diffusion can be qualitatively different from the energy diffusion
in a two-dimensional gas with Lennard-Jones interactions, but the
relaxation behavior of heat has not been studied yet. In previous
studies heat diffusion has constantly been $assumed$ to be normal
in three-dimensional systems, including the three-dimensional gases
with Lennard-Jones interactions \cite{H14,LJ53}. What we have
learned in this work suggests that it is very necessary to check
if this is the case.

\begin{acknowledgments}
This work is supported by the NNSF (Grants No. 10925525,
No. 11275159, and No. 10805036) and SRFDP (Grant No. 20100121110021)
of China.
\end{acknowledgments}


\begin{thebibliography}{99}
\bibitem{note1} We have also checked by nonequilibrium numerical simulations
that in both the FPU model and the gas model, the total heat current and
total energy current are gradually equal to each other after a transient
time.
\end{thebibliography}
\end{document}